          [2001/04/25 1.1 (PWD)]
\documentclass[a4paper,twocolumn]{esapub2005} 
\usepackage{graphicx}
\usepackage{times}
\usepackage{natbib}

\newcommand{\sax}{{\it BeppoSAX} }

\newcommand{\xte}{{\it RXTE} }

\newcommand{\integral}{{\it INTEGRAL} }
\newcommand{\chandra}{{\it Chandra} }

\newcommand{\source}{4U~1705--44 }

\begin{document}
\title{Spectral analysis of the Low-Mass X-ray Binary 4U~1705--44
}
\author{Mariateresa  Fiocchi}
\author{Angela Bazzano}
\author{Pietro Ubertini}
\author{Memmo Federici}
\affil{Istituto di Astrofisica Spaziale e Fisica Cosmica di Roma (INAF). Via Fosso del Cavaliere 100, Roma, I-00133, Italy}
\keywords{accretion, accretion disks -- gamma rays: observations -- radiation mechanisms: non-thermal -- stars: individual: 4U~1705--44 -- stars: neutron -- X-rays: binaries}
\maketitle
\begin{abstract}
\integral\/ and \sax\/ observations of the neutron-star LMXB 4U~1705--44 have been analysed to deeply investigate the spectral state transitions nature.  
Its energy spectrum can be described as the sum of one or two blackbody, a 6.4-keV Fe line and a component due to thermal Comptonization. For the first time in this source, we find a strong signature of Compton reflection, presumably due to illumination of the optically-thick accretion disk by the Comptonized spectrum. Detection of two blackbody component in the soft states could originate in the disk and the neutron-star surface, and the Comptonized component arises from a hot inner flow with the seed photons coming from the disk and/or the neutron-star surface. The spectral transitions are shown to be associated with variations in the accretion rate, which changes in turn the temperature of the Comptonizing electrons and the strength of Compton reflection.
\end{abstract}
\section{Introduction}
\source is a neutron-star low-mass X-ray  binary (LMXB) classified as an
atoll source \citep{has89}. It also shows type-I X-ray bursts \citep{lan87, szt85}  and kHz quasi-periodic oscillations \citep{for98}. The source shows variability on all time scales, from months down to milliseconds \citep{lan87, ber98, for98, liu01}. On long time scales, it displays pronounced luminosity-related X-ray spectral changes between soft and hard states, as illustrated
using data from {\it Rossi X-ray Timing Explorer\/} (\xte) \citep{bar02}.
Using a \chandra/HETGS observation, Di Salvo et al.\ \citep{dis05}
have shown that the iron line is intrinsically broad (FWHM$\sim$1.2 keV), 
confirming the previous values of FWHM$\sim$1.1 keV \citep{whi86, bar02}. 
Compton reflection of X-rays, a process studied in detail in the last few decades
\citep{lig79, whi88, lig88, mag95},
takes place when X-rays
and $\gamma$-rays interact with a cold medium. 
The X-ray spectrum of Compton reflection has a characteristic 
shape resulting from photoelectric absorption at low energies and 
Compton scattering (including its recoil) at higher energies, 
with a broad hump around 30 keV \citep{mat91}. 
\section{Observations and Data Analysis}
Table \ref{jou} gives the log of \integral\/ and \sax\/ observations of the source. \sax\/ observed \source\ twice, in 2000 August and October.
The LECS, MECS and PDS event files and spectra, available from the ASI Scientific Data Center(ASDC), were generated with the Supervised Standard Science Analysis \citep{fio99}. Both LECS and MECS spectra were accumulated in circular regions of 8~arcmin radius. The PDS spectra were obtained with the background rejection method based on fixed rise time thresholds. Publicly available matrices were used for all the instruments. 
Spectral fits were performed in the following energy bands: 0.5--3.8 keV for the LECS, 1.5--10.0 keV for the MECS and 20--150 keV for the PDS.
\begin{small}
\begin{figure}[ht]
\centering
\includegraphics*[angle=-90,scale=0.27]{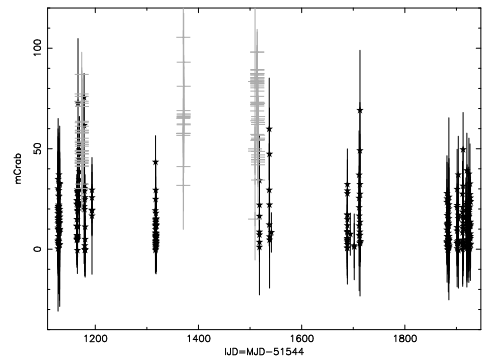}
   \caption{The IBIS light curve in the 40--60 keV energy band. The stars denote data corresponding to relatively steady emission, and the crosses denote data with the average flux $>50\%$ higher than in the average steady level. }
   \label{lci}
   \end{figure}
   \begin{figure}[ht]
\centering
\includegraphics[angle=-90,width=6.8cm]{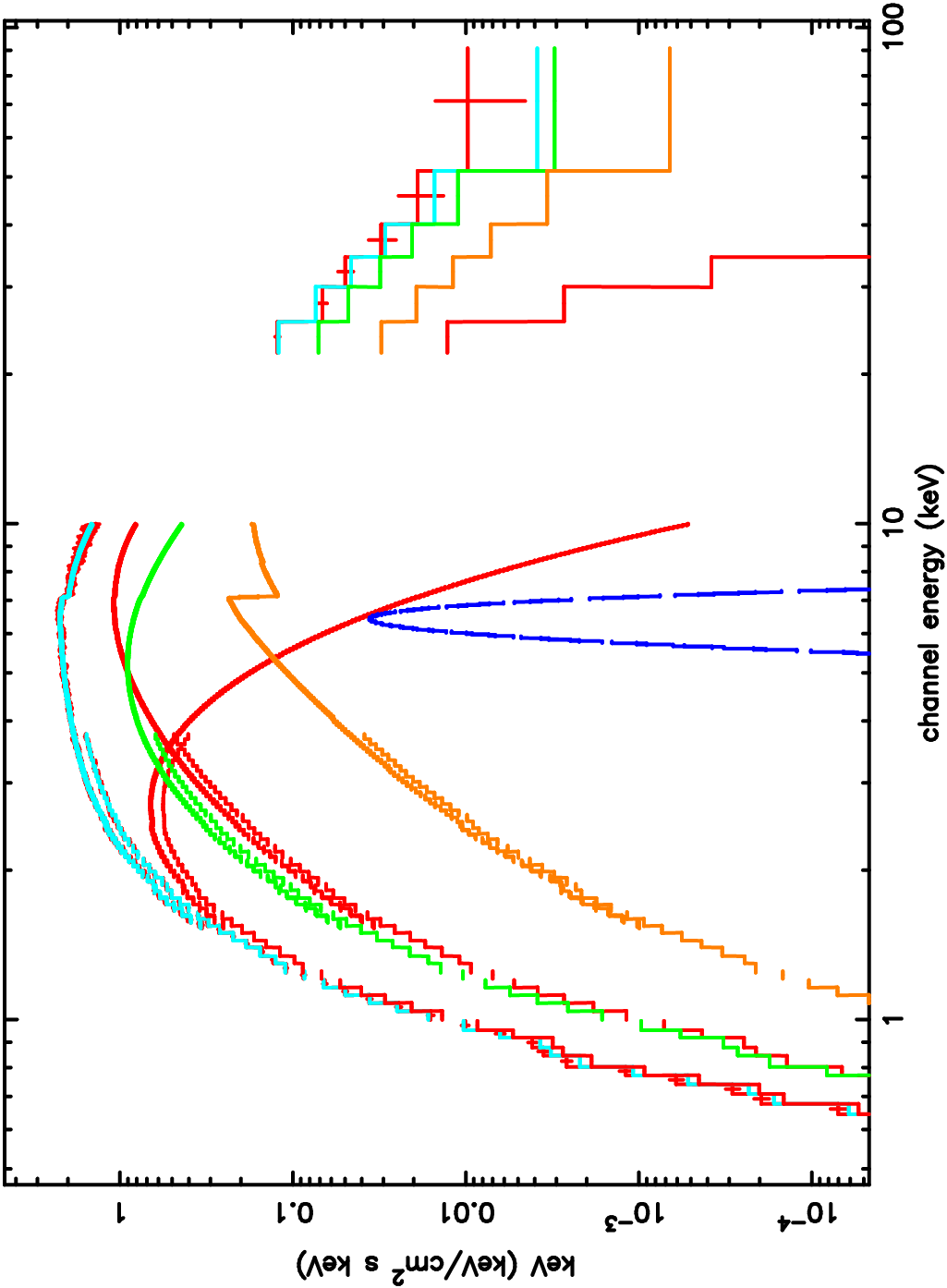}
\includegraphics[angle=-90,width=6.8cm]{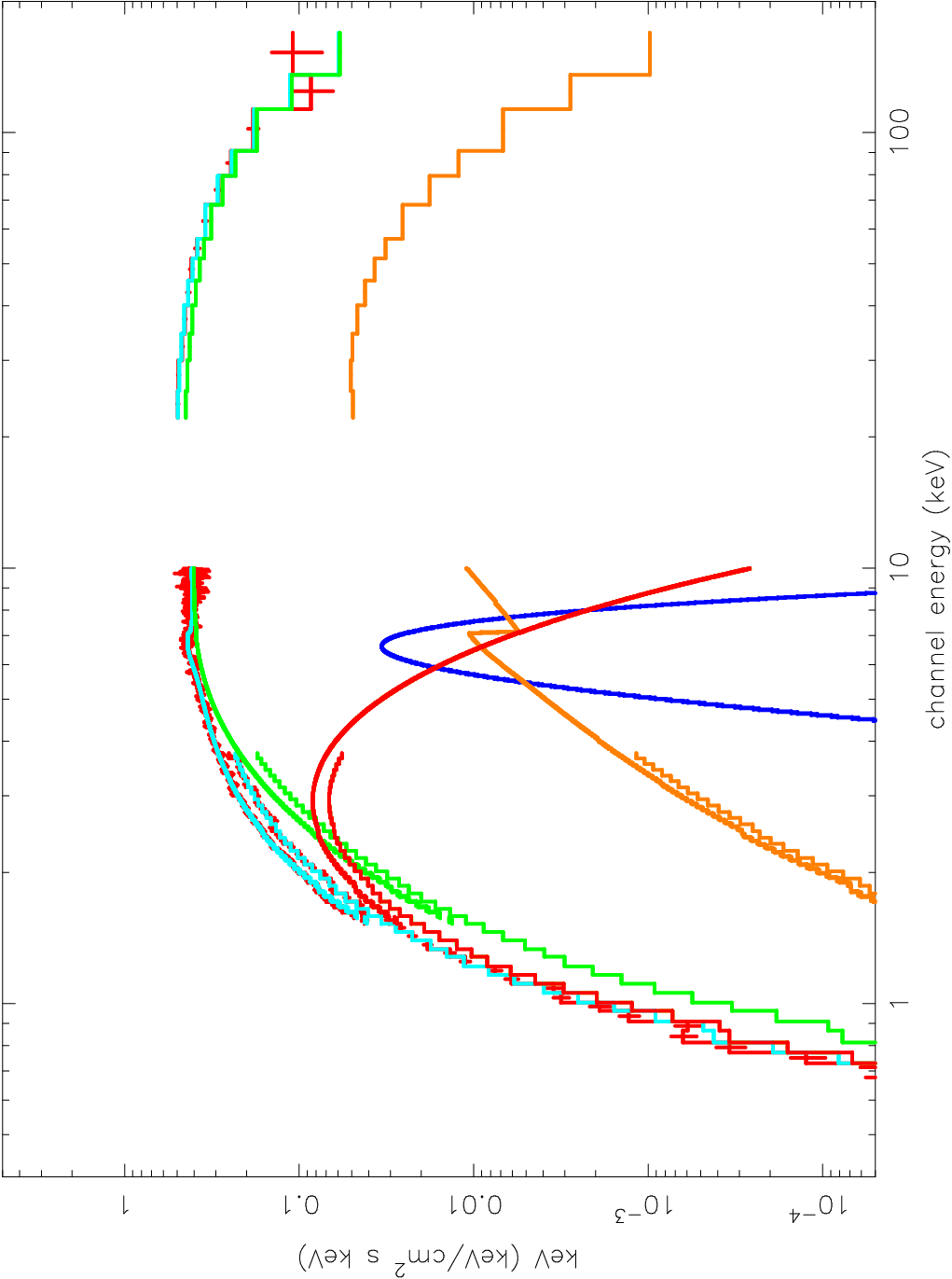}
   \caption{The spectra of epochs 1 (soft state, left) and 2 (hard state, right) observed by \sax\/, shown together with the total model and its components. Left: the blackbody, Comptonization and the Fe line components are shown in red, green and blue, respectively. Right: the blackbody, Comptonization, reflection and Fe line are shown in red, green, orange and blue respectively.
}
   \label{sax1}
   \end{figure}
\end{small}
The analyzed \integral\/ \citep{win03} data consist of all public observations in which 4U 1705--44 was within the field of view (FOV) of high-energy detectors. The observations are divided into uninterrupted 2000-s intervals, the so-called science windows (SCW), within which the light curves and spectra are extracted individually. Broad-band spectra, $\sim$5--150 keV, are obtained using data from the high-energy instruments, JEM-X \citep{lun03} and IBIS \citep{ube03}. The IBIS and JEM-X data have been processed using the Off-line Scientific Analysis (OSA v.\ 5.0) software released by the \integral\/ Science Data Centre (ISDC, \cite{cou03}). 
Data from the Fully Coded field of view only for both instrument have been used
($4.5^o$ and $2.5^o$ for IBIS and JEM-X, respectively). Spectra were extracted only if the ratio signal-to-noise was $>10$.
\section{Spectral Analysis Results}
\begin{table*}
\caption{The log of \sax\/ and \integral\/ observations.}
\label{jou}
\centering
\footnotesize
\vspace{0.1cm}
\begin{tabular}{lccccccc}
\hline
\multicolumn{8}{c}{\bf{BeppoSAX}}\\
& {Start Date}&&{Exposure time [ks]}&&&{Count rate [s$^{-1}$]$^{a}$}&\\
&&LECS&MECS&PDS&LECS&MECS&PDS\\
&year-month-day& & & &[0.5--3.8 keV]&[1.5--10 keV]&[20--100 keV]\\
& & & & & & &\\
{$1^{\rm st}$ epoch} &2000-08-20 &21 &44 &20 &$3.88\pm 0.01$  &$27.25\pm 0.03$ &$0.90\pm0.04$\\
{$2^{\rm nd}$ epoch} &2000-10-03 &16 &48 &20 &$0.48\pm 0.01$  &$4.46\pm 0.02$  &$7.86\pm0.05$\\
\hline
\multicolumn{8}{c}{\bf{INTEGRAL}}\\
&&{Start Date}&\multicolumn{2}{r}{Exposure time [ks]}&\multicolumn{2}{c}{Count rate [s$^{-1}$]}&\\
&& &JEM-X&IBIS&JEM-X&IBIS&\\
&&year-month-day& & &[4.5--15keV]&[20--150 keV]&\\
& & & & & &&\\
&{$3^{\rm rd}$ epoch}  &2003-02-02  &106  &394   &$11.63\pm 0.04$ &$1.60\pm0.05$&\\
&{$4^{\rm th}$ epoch} &2003-08-09  &17  &93 &$8.6\pm 0.1$ &$12.6\pm 0.1$&\\
\hline
\end{tabular}\\
{$^{a}$ The MECS count rates correspond to the sum of the MECS2 and MECS3 units.}
\end{table*}
To determine the parameters driving the spectral state changes, we fit 
the data 
with several physical models.
 Each time a new component is added to the model, the F-test 
is performed. We assume that an F-test probability of $>95\%$ implies a 
significative improvement of the fit. 
We use XSPEC v.\ 11.3.1.
Fig.\ \ref{lci} shows IBIS light curve in the 40--60 keV energy band,
where we expect main spectral changes for atoll sources. As it can be seen the source
flux has been observed to increase more than $50\%$ on three occasions, during MJD
 52712--52718, 52914--52916 and 53044--53060, as showed in Fig.\ \ref{lci} as crosses.
However, we find substantial spectral changes only during the last two intervals.\\
We study the spectral behavior separately in four epochs consisting of the following
 data:
\begin{itemize}
\item
{$1^{\rm st}$ epoch}. The \sax\/ observation performed on MJD 51777 
during which the source was in the soft/high state.
\item
{$2^{\rm rd}$ epoch}. The \sax\/ observation performed on MJD 51820 
during which the source was in the hard/low state.
\item
{$3^{\rm rd}$ epoch}. The \integral\/ JEM-X and IBIS observations between MJD 52671
to 53471 but excluding the observations of MJD 52914--52916 and 53044--53060.
\item
{$4^{\rm th}$ epoch}. The JEM-X and IBIS observations of MJD 52914--52916 and 53044-
-53060.
\end{itemize}
\begin{table*}[]
\scriptsize
\centering
\caption{The fit results.
'f' denotes a fixed parameter. }
\label{fit}
\vspace{0.1cm}
\begin{tabular}{c@{\ \ }c@{\ \ }c@{\ \ }c@{\ \ }c@{\ \ }c@{\ \ }c@{\ \ }c@{\ \ }c@{\
 \ }c@{\ \ }c@{\ \ }c@{\ }c@{\ }c}
\hline
&{$N_{\rm H}$} &{kT$_{\rm BB1}$} &{kT$_{\rm BB2}$} & $T_{0}$  &{kT$_{\rm e}$} &$\tau
$ &$\Omega/2\pi$ &$\sigma_{\rm Fe}$& EW  &$R_{BB1}$ &$R_{BB2}$& $n_{Compt
t}$ & $\chi^2$/d.o.f. \\
&$10^{22}$cm$^{-2}$ &keV &keV&keV &keV &&& keV &eV  &     km&km      &$10^{-2}$
&\\
\hline
& & & & & & & & & &&&\\
1&$1.37\pm0.03$  &$0.58^{+0.01}_{-0.01}$&$1.75\pm0.05$  &$1.06\pm0.05$ & $14.9_{-0.5
}^{+1.0}$  &$0.4\pm0.1$&$1.1\pm0.2$ & $<0.3$ & $14\pm6$  &$31.0\pm0.5$ &
$3.7\pm0.4$& $2.6\pm0.3$ & 542/421\\
& & & & & &  & & &&&&\\
2 &$1.2_{-0.1}^{+0.1}$ &$0.65_{-0.06}^{+0.05}$&&$1.14_{-0.08}^{+0.04}$   &$22_{-1}^{
+4}$ &$1.5_{-0.6}^{+0.3}$        &$0.2_{-0.1}^{+0.2}$  & $0.8^{+0.3}_{-0.3}$ & $189_
{-39}^{+20}$  & $8.2_{-0.1}^{+0.1}$& &$0.7\pm0.1$ & 500/413\\
& & & & & & & & & &&&\\
3 & 1.3f&0.57f&$2.4\pm0.1$
&1.07f &$16\pm1$ &$0.3\pm0.1$ &$<0.6$ & & &$<33$& $1.5\pm0.2$ &$4.3\pm0.3$&116/1
09\\
& & & & & & & & & &&&\\
4 &1.3f &0.65f& &1.14f & $49\pm3$  &$0.3\pm0.1$&$1.4_{-0.3}^{+0.5}$  & & &$<17$&
&$0.5\pm0.2$ &118/116\\
& & & & & & & & & &&&\\
\hline
\end{tabular}\\
\end{table*}
\begin{small}
 \begin{figure}[h]                                                                 \centering
  \includegraphics[angle=-90,width=6.8cm]{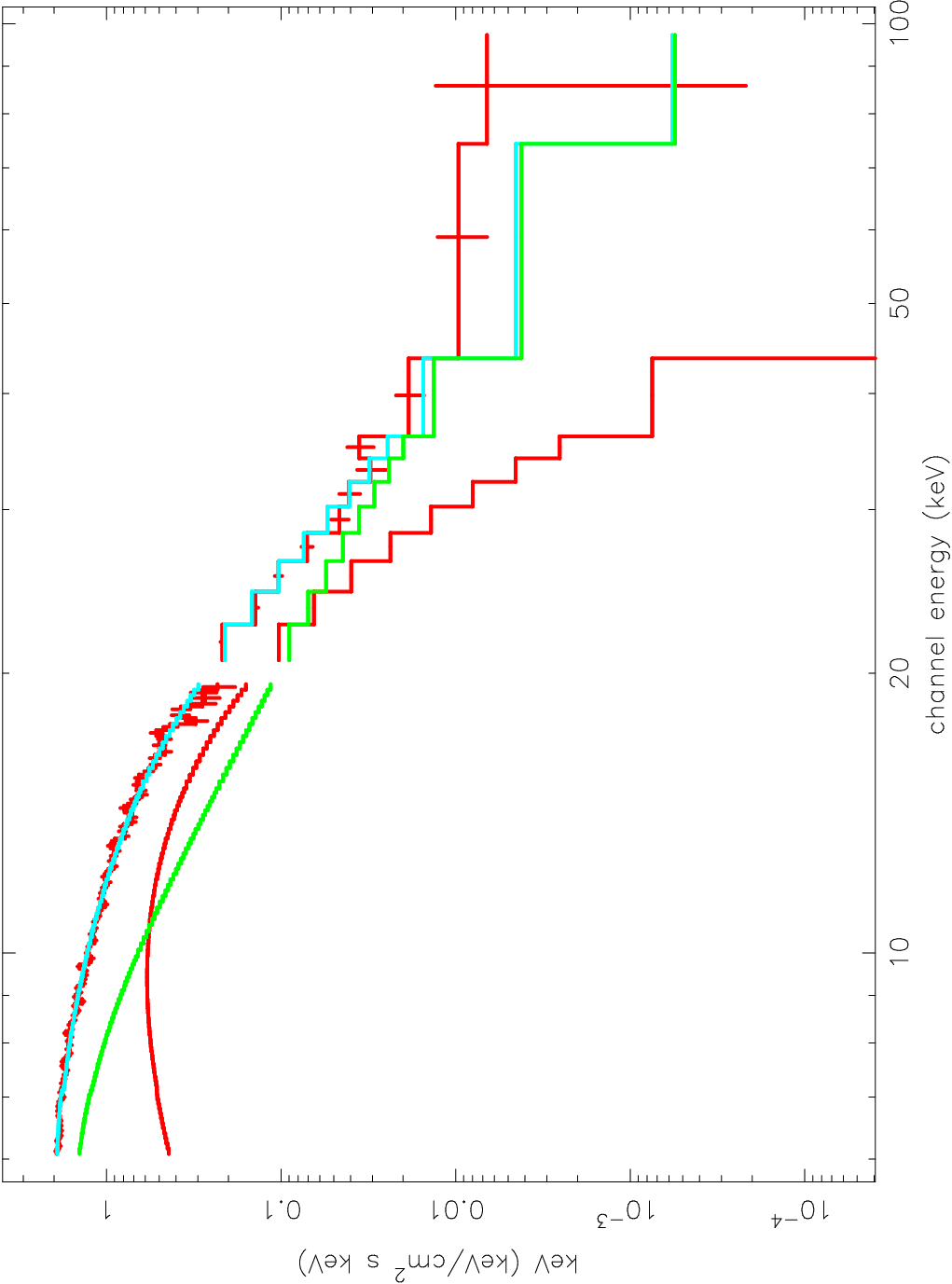}                                    \includegraphics[angle=-90,width=6.8cm]{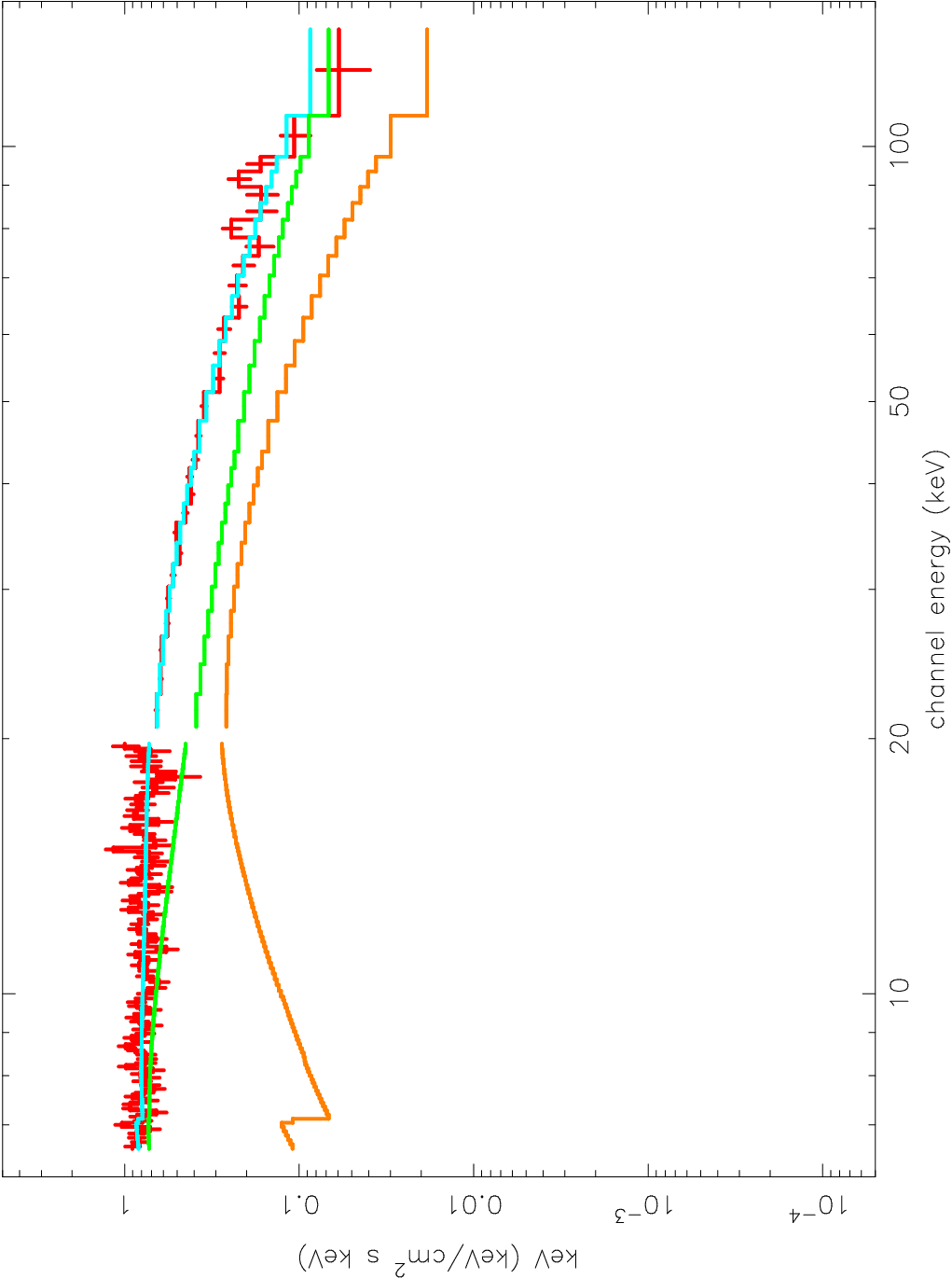}                                    \caption{The spectra of epochs 3 (soft state, left) and 4 (hard state, right) observed by \integral, shown together with the total model and its components. Left: the blackbody and Comptonization components are shown in red and green, respectively. Right: Comptonization and the reflection component is shown in green and orange, respectively.}                                                                             \label{int1}                                                                        \end{figure}
\end{small}
The simplest model which provides a reasonable fit to the \sax\/ data consists of th
e sum of a thermal Comptonization component, modeled in XSPEC by
{\scriptsize{COMPTT}} (\cite{tit94}; spherical geometry was assumed),
a soft component modeled by a single temperature blackbody, and a Gaussian Fe line at 6.4 keV with the width, $\sigma_{\rm Fe}$. 
This spectrum is absorbed by a column density, $N_{\rm H}$, 
consistent with that given by
Langmeier et al.\ \citep{lan87}.
Initially, the Fe line centroid energy was let free; however, our fits found it to be always consistent with the neutral-Fe value of 6.4 keV, and hence it has been fixed at that value hereafter.
Adding Compton reflection to the fit, using the XSPEC model {\scriptsize{REFLECT}}
\citep{mag95}, resulted in a substantial fit improvement for epoch 2,
reducing $\chi^2/$d.o.f.\ from $510/414$ to $500/413$. The inclination angle, $i$, of the reflecting medium has been kept at $63^o$ ($\cos i=0.45$), which is consistent with the allowed range found by Di Salvo et al.\ \citep{dis05} of $i=55$--$84^o$. The strength of reflection is measured by the solid angle, $\Omega$, subtended by the reflection.
The black body temperatures are $T_{BBsoft}=0.58keV$
and $T_{BBhard}=0.65keV$ for the soft and hard state respectively.\\
The \integral\/ data have been fitted with
a model
similar to that used for the \sax\/ data.
However, since both the energy resolution of JEM-X, $\Delta E/E \sim 1$ (FWHM) at 6.4
keV is not adequate to study the Fe line properties and the usable
JEM-X data are for $>~5$ keV only,
we do not include the line in the present model.
The seed soft temperature and the column density are now
frozen at their respective values obtained from the \sax\/ data.
Similary to the case of the
\sax\/ data, we add the Compton reflection to the fit, using the XSPEC
model {\scriptsize{REFLECT}} with $i=63^o$.
Since the \integral\/ energy range could not be
reliable in determining the black body temperature at low energy,
we fix it at the value obtained from the
hard state \sax\/ data.
This procedure is not good for the epoch 3:
fixing it at the value obtained
from the \sax\/ soft state data, the fit gives a $\chi^2$ not acceptable
($\chi^2/_{rid}\sim2$), a higher temperature of soft component is required by data
($T_{BB}\sim2.5 keV$).
For this reason, we fit both \sax\/ and \integral\/ soft states with two
black body components,
reducing $\chi^{2}/$d.o.f. from 578/423 to 542/421 for the epoch 1 and
from 195/109 to
120/111 for the epoch 3.
Spectral fit results are given in Table \ref{fit}
and the spectra are shown in Fig.\ \ref{sax1} and \ref{int1}
for \sax\/ and \integral, respectively.
In the soft/high state
the extrapolated model luminosities are
$L_{0.1-200\,\rm keV} \simeq 4.8 \times 10^{37}$ erg s$^{-1}$
and $\simeq 3.2 \times 10^{37}$ erg s$^{-1}$, for the epochs 1 and 3,
respectively, at the distance of 7.4 kpc \citep{hab95}.
Most of the source flux, $\sim 80 \%$,
is radiated below 20 keV. 
In the hard/low state
the luminosities are $L_{0.1-200\,\rm keV} \simeq 1.5 \times 10^{37}$ erg s$^{-1}$
and $\simeq 2.1 \times 10^{37}$ erg s$^{-1}$, for the epochs 2 and 4, respectively.
They are both lower than those in the soft state,
in agreement with the usual ranking of the luminosity of these two states,
in particular in atoll sources \citep{has89, van00, gie02}.
The electron temperatures are now substantially higher than in the soft state,
$kT_{\rm e}\sim 20$--50 keV, and the Comptonization component
extends above $\sim 100$ keV.
The seed-photon temperatures are consistent
with values typical for neutron-star LMXBs,
$kT_{0}\sim 0.3$--1.5 keV \citep{oos01}.
The black body temperature $kT_{\rm BB2}$ is high during our soft states,
similar high temperatures have been found by Barret \& Olive \citep{bar02}
in the soft states \xte\/ observations. 
No significant variations of $kT_{0}$ or $N_{\rm H}$ have been detected,
with our values consistent with previous ones \citep{bar02, lan87}.
\section{Conclusions}
\label{discussion}
According to our present understanding, the low-energy, black-body component in the soft states originates at both the neutron star surface and the surface of an optically-thick accretion disk. The Comptonization component may arise from a corona above the disk and/or
between the disk and the stellar surface.
The obtained values of the electron temperature are typical for neutron-star LMXBs,
 and are lower than those seen in the hard state of black-hole binaries
($kT_e\sim 100$ keV, e.g., \cite{gie97, dis01, zdz04}),
confirming the idea the electron temperature could be used as a criterion
distinguishing between black-hole and neutron-star binaries
\citep{tav97, zdz98, bar00}.
In the two \sax\/ data sets we have detected a
iron line emission which we interpret as K-shell fluorescent
emission of the iron in low ionization states.
For the first time in this source, we have detected strong signatures of
Compton reflection,
earlier detected in same 
of neutron-star LMXBs, e.g., in GS 1826--238 and SLX 1735--269 \citep{bar00},
4U 0614+09 \citep{pir00}, GX 1+4 \citep{rea05}.
The strong iron line detected in
the 2 epoch when the reflection is weak indicates iron
line intensity and reflection component do not vary together and
geometrical factor could be responsible of this behavior.\\
The spectral transitions are accompanied by changes in luminosity,
indicating they are driven by variability of the accretion rate,
being lower in the hard state and higher in the soft state.
\begin{small}

\end{small}
\section*{Acknowledgments}
We acknowledge the ASI financial/programmatic support via contracts ASI-IR
 046/04.
\end{document}